\documentstyle [twocolumn,epsf]{mn}
\newcommand{\mincir}{\raise
-3.truept\hbox{\rlap{\hbox{$\sim$}}\raise4.truept\hbox{$<$}\ }}
\newcommand{\magcir}{\raise
-3.truept\hbox{\rlap{\hbox{$\sim$}}\raise4.truept\hbox{$>$}\ }}
\newcommand{\minmag}{\raise
-3.truept\hbox{\rlap{\hbox{$<$}}\raise5.truept\hbox{$<$}\ }}
\newcommand{\be}{\begin{equation}}
\newcommand{\ee}{\end{equation}}
\newcommand{\ba}{\begin{eqnarray}}
\newcommand{\ea}{\end{eqnarray}}
\newcommand{\brr}{\begin{array}}
\newcommand{\err}{\end{array}}
\newcommand{\bc}{\begin{center}}
\newcommand{\ec}{\end{center}}

\title[Cluster substructure]{Searching for cluster substructure using 
APM and ROSAT data.}
\author[Kolokotronis et al.]{V. Kolokotronis$^{1}$, S. Basilakos$^{2,1,3}$,
M. Plionis$^{1}$, I. Georgantopoulos$^{1}$. \\
\vspace{0.1cm}
$^1$ Institute of Astronomy \& Astrophysics, National Observatory of Athens, 
Lofos Nimfon, Thesio, 18110 Athens, Greece \\
$^2$ Astrophysics Group, Imperial College London, Blackett Laboratory, 
Prince Consort Road, London SW7 2BW, UK\\
$^3$ Physics Department, University of Athens, Panepistimiopolis, Greece \\
}

\begin{document}

\maketitle

\begin{abstract}
We present a detailed study of the morphological features 
of 22 rich galaxy clusters. Our sample is constructed
from a cross-correlation
of optical (Abell+APM) data with X--ray (0.1 - 2.4) keV ROSAT pointed 
observations. 
We systematically compare cluster images and morphological parameters 
in an attempt to reliably identify possible substructure in both
optical and the X--ray images. To this end, we compute various
moments of the optical and X--ray surface-brightness distribution such as
the ellipticities, center-of-mass shifts and ellipsoidal orientations. 
We assess the significance of our results using Monte Carlo simulations.
We find significant correlations between the optical and X--ray morphological
parameters, indicating that in both parts of the spectrum it is possible to
identify correctly the dynamical state of a cluster.
Most of our clusters (17/22) have a good 1-to-1 correspondence between
the optical and the X--ray images and about 10 appear to have strong
indications of substructure. This corresponds to a minimum percentage of 
order $\sim 45 \%$, which is in very good accordance
with other similar analyses. Finally, 5 out of 22 systems 
($\sim 22 \%$) seem to have distinct subclumps in the optical which are not
verified in the X--ray images, and thus are suspect of being due to
optical projection effects. These results will serve as a useful guide 
in interpreting subsequent analyses of large optical cluster catalogues.

{\bf Keywords:} galaxies: clusters: general - large-scale structure of 
universe -  X--ray: galaxy clusters
\end{abstract}

\vspace{0.2cm}

\section{Introduction}

Galaxy clusters occupy a special position in the hierarchy of cosmic 
structure in many respects. Being the largest physical laboratories in the 
universe, they appear to be ideal tools for studying large-scale structure, 
testing theories of structure formation and extracting invaluable 
cosmological information, especially regarding the Hubble parameter and 
the value of $\Omega_{\circ}$ (cf. B${\rm \ddot{o}}$hringer 1995; 
West, Jones \& Forman 1995; Buote 1998; Schindler 1999).

One of the most significant properties of galaxy clusters is the relation
between their dynamical state and the underlying cosmology. 
In an open universe, clustering effectively freezes at high redshifts 
($z\simeq\;\Omega_{\circ}^{-1}-1$) and 
clusters today should be more relaxed with 
weak or no indications of substructure. 
Instead, in a critical density model, such systems continue to form even today
and should appear to be dynamically active.
Even, in a small but
well-controlled (from any kind of selection or other biases) cluster sample, 
the percentage and morphologies of disordered and perturbed 
objects could lead to constraints on the $\Omega_{\circ}$ and $\Lambda$ parameters, 
especially if combined with N-body/gasdynamic numerical simulations spanning
different dark matter (DM) scenarios (cf. Richstone, Loeb \&
Turner 1992 hereafter RLT92; Evrard et al. 1993; Lacey \& Cole 1993).

The above pioneering works were the first to set some limits on 
$\Omega_{\circ}$ using the rate of cluster formation for various 
cosmologies and led the way to a plethora of research works towards this 
direction (see Figure 2 of RLT92). Since then, a large number of relevant 
analyses have been devoted to this study and an accordingly varying and 
large number of optical and X--ray cluster compilations have been utilised 
to this aim. All the methods employed in each case are quite different and 
all the related studies find an appreciable percentage of dynamically 
active galaxy clusters (see Forman \& Jones 1990; B${\rm \ddot{o}}$hringer 
1995; West 1995; Thomas et al. 1998 for good reviews of the subject). 
However, these studies do disagree on the precise number of clusters exhibiting
significant dynamical activity,
which varies between 30\% and 80\% of the total number of clusters studied,
and which seems also to depend on the techniques employed in each 
analysis.

We quote only some of the most recent studies together with the 
sample used and the preference of the method given in each case. 
Geller \& Beers (1982) analysed the iso-intensity contour maps of 65 
optical clusters maps. Forman et al. (1981) studied 4 Einstein
clusters. Dressler \& Shectman
(1988) studied velocity dispersion diagnostics for 15 optical 
clusters. Rhee, van Haarlem \& Katgert (1991) and Solanes, 
Salvador-Sole \& Gonzalez-Casado (1999) utilised 107 and 67 ENACS clusters 
respectively and a variety of 2D and 3D statistical tests to quantify the 
significance of substructure. Similarly, Dutta 
(1995), Crone, Evrard \& Richstone (1996 hereafter CER96), Pinkney et al. 
(1996), Thomas et al. (1998 hereafter T98) and Jones \& Forman (1999 
hereafter JF99; 208 ROSAT X--ray clusters) proposed a variety of 
substructure tests depending on the level of information available (1D, 2D, 
3D) and applied these to different N-body cluster data using several DM 
models. CER96 and T98 have explicitly argued that variations in the 
cluster center-of-mass as a function of distance or density (overdensity) 
threshold is one of the best possible substructure measures (see also West 
\& Bothun 1990). The latter has previously been adopted by Mohr, Fabricant \& 
Geller (1993; 5 X--ray clusters), Mohr et al. (1995; 65 Einstein clusters) 
and Rizza et al. (1998; 11 ROSAT HRI distant clusters). Evrard et al. 
(1993) and Mohr et al. (1993; 1995) were the first to make extensive 
use of the surface-brightness moments of the X--ray cluster
distribution (eg. center-of-mass shifts). 
Gomez et al. (1997; 9 Abell clusters) have 
also employed the same methods and 
claimed that variations in the cluster ellipticity and orientation as a 
function of distance from the cluster center (isophotal twisting) should be 
considered as one of the prime substructure diagnostics. Bird (1994) in her 
sample of 25 Abell cluster has utilised galaxy peculiar velocities to 
locate and identify subclumps in the galaxy distribution. Kriessler \& Beers 
(1997; 56 Abell and other clusters) have searched for cluster substructure 
signatures using the innovative KMM algorithm on the surface-density galaxy 
maps and quantified their findings using N-body simulations. 
Furthermore, Buote \& Tsai (1995; 1996 hereafter BT95; BT96) and 
Valdarnini, Ghizzardi \& Bonometto (1999) used the 2D gravitational 
potential moments (power ratio method) to characterise the dynamical state 
of clusters. Serna \& Gerbal (1996; 2 Abell 
clusters) have developed the so-called hierarchical method to define and 
identify substructure in Abell clusters, while
Slezak et al. (1994; 11 X--ray clusters) and Lazzati et al. (1998; 2 X--ray 
clusters) have adopted wavelet transform techniques in similar attempts.

As it is evident from all the above there is neither agreement on the methods 
utilised nor on the exact frequency of perturbed clusters. It seems that 
identifying significant dynamical activity within galaxy clusters, in close 
relation to the underlying cosmology and the density parameter, still remains 
an open issue. Not only do we need a large, statistically complete, cluster 
sample but an objective and reliable definition of what substructure is, upon 
which our study should be based. 

The large majority of the analyses carried out so far, have made use 
of either optical (Abell)
or X--ray (ROSAT, Einstein) cluster data. However, Mohr et al. (1993) and 
Rizza et al. (1998) have investigated cluster substructure using in a 
complementary fashion optical and X--ray data. In the present work, we 
extend this approach using a sample of 22 galaxy clusters for which we have 
data both in the optical and the X--ray part of the spectrum (APM and ROSAT 
respectively). Our sample size is twice as large as that of Rizza et al. 
(1998) and more than four times larger than that of Mohr et al. (1993). The 
advantage of using X--ray data is that the X--ray emission is proportional 
to the square of the gas density (rather than just density in the optical) 
and emanates mostly from the central cluster region, a fact which minimises 
projection effects (cf. Sarazin 1988; Schindler 1998). The advantage of 
using optical data is the shear size of the available cluster 
catalogues and thus the statistical significance of the emanating results. 
Subsequently, we are not 
only interested in comparing optical to X--ray cluster data regarding the 
various substructure tools, but to calculate and calibrate different biases 
using the superior X--ray data with which we could measure and test easily 
similar substructure diagnostics in a large, solely optical dataset as well. 
Therefore it is of great importance to address the following two questions, 
which we attempt to do in our present study:

\begin{itemize}
\item Is substructure in the optical also corroborated by the X--ray 
observations and in what percentage?
\item  What is the confirmed percentage of systems depicting significant 
indications of internal activity?
\end{itemize}

To this aim we compare optical to X--ray morphological cluster parameters 
(position angles, ellipticities, centroid shifts and group statistics) in an 
attempt to classify objects according to their dynamical state and we 
compute the relative frequency of substructure.

We proceed by presenting the optical and X--ray datasets in section 2. In 
section 3, we describe our methods of analysis, in section 4 we present a 
comparison of the optical and X--ray cluster images while in section 5 we 
present the results of our substructure analysis. Finally, we draw our 
conclusions in section 6. Also, there is an extensive appendix at the end of 
this paper, where morphological and dynamical information for each individual 
cluster can be found as well as a comparison with other similar works on the 
common clusters.

\section{The data}

The present dataset follows from a double cross-correlation between very rich 
ACO (Abell, Corwin \& Olowin 1989) clusters ($\rm R\geq\;$1,2,3) with the APM 
cluster catalogue (Dalton et al. 1997 and references therein) and the X--ray 
(0.1 - 2.4) keV ROSAT pointed observations archive. The first correlation 
results in 329 common optical galaxy clusters (ACO/APM) in the southern sky 
($b\leq 40^{\circ}, \delta\leq -17^{\circ}$), while the second correlation 
results in 27 common clusters. Due to problematic regions of the APM 
catalogue, low signal to noise X--ray observations, contamination by known 
foreground or background objects and even double entries we exclude 5 
clusters (A4038, A122, A3264, A3049 and A2462) reducing our cluster sample 
to 22 systems.

Furthermore, the ROSAT (Tr${\rm \ddot{u}}$mper 1983) data we have used come 
from both the PSPC (Positional Sensitive Proportional Counter; Pfefferman 
\& Briel 1986), and the HRI (High Resolution Imager) detectors, both 
operating in the (0.1 - 2.4) keV band. The HRI has an excellent spatial 
resolution (FWHM$\;\sim$ 5 arcsec) but no spectral resolution. The PSPC has 
an energy resolution of 0.5 keV at 1 keV but poorer spatial resolution (30 
arcsec FWHM) compared to HRI. We note that the PSPC is quite efficient in 
detecting extended emission, in contrast with the HRI which has a lower 
sensitivity due to its higher background. 

The cluster redshifts of our sample span the range $0.04 \mincir 
z \mincir 0.13$ with $\langle z \rangle \sim 0.074$ and median $\sim 0.069$. 
For the needs of our analysis we transform cluster redshifts to cluster
distances 
using the well-known angular-diameter distance formula with $q_{\circ}=0.5$ 
(ie., critical density universe) as:

\begin{equation}\label{eq:dist}
r=\frac{2c}{H_{\circ}}\;(1+z)^{-1/2}\;[1-(1+z)^{-1/2}]\;\;,
\end{equation}
with $H_{\circ}=100 \;h$ km $s^{-1}$ Mpc$^{-1}$.
In Table 1 we give all the relevant details for the present cluster sample. 

\begin{table}
\caption[]{Details of our cluster sample: APM and ACO index numbers, 
equatorial coordinates (1950) of cluster centers in degrees, heliocentric 
redshift, richness class, type of ROSAT observation and exposure time 
(in seconds). All cluster redshifts are taken from Katgert et al. (1996) and 
Mazure et al. (1996), apart from those of A2804 (Dalton et al. 1994), 
A2811 (Collins et al. 1995) and A2580 (White, Jones \& Forman 1997). For A3144 
and A2580 we have used $m_{10}$ estimated redshifts.}
\tabcolsep 3pt
\begin{tabular}{cccccccr}
\hline
APM & ACO & $\alpha$ & $\delta$ & $z$ & ${\rm R}$ & Type & $t_{\rm exp}$ 
\\ \\ \hline  \hline 
 5  & 2717 & 0.1083  & -36.3113 & 0.0498  & 1 & PSPC & 10000 \\
15  & 2734 & 2.1783  & -29.1011 & 0.0620  & 1 & PSPC &  5000 \\
99  & 2804 & 9.2962  & -29.1639 & 0.1080  & 1 & PSPC &  7500 \\
104 & 2811 & 9.9729  & -28.8475 & 0.1090  & 1 & HRI  &  9350 \\
138 & 0133 & 15.0516 & -22.2228 & 0.0570  & 1 & HRI  & 12500 \\
204 & 2933 & 24.6966 & -54.8097 & 0.0930  & 1 & PSPC &  7900 \\
347 & 3093 & 47.4212 & -47.6186 & 0.0830  & 2 & PSPC &  8100 \\
373 & 3111 & 49.0317 & -45.8886 & 0.0780  & 1 & PSPC &  7600 \\
374 & 3112 & 49.0337 & -44.4247 & 0.0750  & 2 & HRI  &  9600 \\
403 & 3128 & 52.2929 & -52.7231 & 0.0600  & 3 & HRI  &  6500 \\
427 & 3144 & 54.0304 & -55.2681 & 0.0450  & 1 & PSPC &  2921 \\
434 & 3158 & 55.3371 & -53.7884 & 0.0590  & 2 & PSPC &  3050 \\
484 & 3223 & 61.5212 & -31.0419 & 0.0601  & 2 & PSPC &  7700 \\
510 & 3266 & 67.6329 & -61.5120 & 0.0589  & 2 & HRI  &  8202 \\
518 & 0500 & 69.1946 & -22.1964 & 0.0670  & 1 & PSPC & 18400 \\
533 & 0514 & 71.5001 & -20.5692 & 0.0730  & 1 & PSPC & 18105 \\
560 & 3301 & 74.8496 & -38.8044 & 0.0540  & 3 & PSPC &  8886 \\
725 & 2384 & 327.385 & -19.8139 & 0.0943  & 1 & HRI  & 26144 \\
806 & 3897 & 339.129 & -17.6010 & 0.0733  & 1 & PSPC &  4300 \\
822 & 3921 & 341.5895& -64.6525 & 0.0940  & 2 & PSPC & 12000 \\
888 & 2580 & 349.6854& -23.4619 & 0.1297  & 1 & HRI  & 17663 \\
938 & 4059 & 358.6346& -34.8300 & 0.0488  & 1 & HRI  &  6320 \\ \hline  
\end{tabular}
\end{table}

\section{The methodology}

In this section we provide an account of the techniques
used to define the cluster morphological parameters as well as the
substructure measures. We first present how to ${\it process}$ the cluster 
images and reduce the noise from point-like sources.

\subsection{Processing the cluster images}
Our X--ray cluster images are retrieved from the ROSAT pointed 
observations archive and we have used the image processing package 
XIMAGE, which is designed to display and reduce the available data. 
Each X--ray image is embedded in a $512 \times 512$ grid, each cell of 
which has a size of 15 arcseconds. We subtract all known point-like 
sources around the clusters (cf. West 1995; JF99) and replace them 
with the average background counts. In doing so, we have mostly 
encountered radio point-like sources (see Appendix for details). 

Our optical data consist of all APM galaxies that fall within a radius of 
$1.8 \;h^{-1}$ Mpc from each optical APM cluster center. We then transform 
the galaxy and X--ray grid-cell angular coordinates into physical units, 
centering the X--ray data on the optical cluster centers. In order to 
construct a common comparison base, we create a continuous density field for 
both optical and X--ray data by using a Gaussian Kernel and the same 
smoothing length ($R_{\rm sm}$). To this end we utilise a $N \times N$ 
grid, where the typical size of each grid cell is $\sim 0.065 \; h^{-1}$ Mpc. 
In order to take into account the reduction of the 
number of galaxy cluster members as a function of distance (due to the APM magnitude 
limit), and thus the corresponding increase of discreteness effects, we have 
investigated, using Monte-Carlo cluster simulations, the necessary
increase in size of $N$ and $R_{\rm sm}$, as a function of distance, 
in order to minimise such effects and optimize the performance 
of our procedure (for details see Basilakos, Plionis \& Maddox 2000; 
hereafter BPM). 

\subsection{Cluster Shape Parameters}

We compute the optical and X--ray cluster shape parameters utilising the 
method of moments of inertia (cf. Carter \& Metcalfe 1980; Plionis, Barrow 
\& Frenk 1991; BPM). We can then write the moments as follows:

\begin{equation}\label{eq:i1}
I_{\rm 11}=\sum_{i=1}^{N} \rho_{\rm i}\;(r_{\rm i}^{2}-x_{\rm i}^{2}) 
\end{equation}

\begin{equation}\label{eq:i2}
I_{\rm 22}=\sum_{i=1}^{N} \rho_{\rm i}\;(r_{\rm i}^{2}-y_{\rm i}^{2})
\end{equation}

\begin{equation}\label{eq:i3}
I_{\rm 12}={\it I}_{\rm 21}=-\;\sum_{i=1}^{N} \rho_{\rm i}\;x_{\rm i}\;
y_{\rm i}\;,
\end{equation}

\noindent 
where $\rho_{\rm i}$ is the cell density, $x_{\rm i}$, $y_{\rm i}$ are the 
Cartesian coordinates of the grid cells and $r_{\rm i} = 
\sqrt{x_{\rm i}^2 + y_{\rm i}^2}$. If we now diagonalise the inertia tensor 
$\det\;(I_{\rm ij}-\lambda^{2}\,M_{2})=0$ ($M_{2}$ being the $2\,\times\,2$ 
unit matrix), we can obtain the eigenvalues $\lambda_{1}, \lambda_{2}$, from 
which the ellipticity of each object can be estimated as 
$\epsilon\,=\,1-\lambda_{2}/\lambda_{1}$, with ${\lambda_{1}}>{\lambda_{2}}$. 
The eigenvectors, corresponding to these eigenvalues, provide us with the 
cluster orientations. The major axis orientation with respect to the North, 
in the anticlockwise direction, is the so-called cluster position angle 
($\theta$ hereafter).

The shape parameters are estimated using all cells that
have densities above three thresholds. These are defined as the average 
density of all cells that fall within a chosen radius. 
The three radii used are $r_{\rm \rho}=0.3, 0.45$ and $0.6 \;h^{-1}$ Mpc. 
The choice of such a step size is not arbitrary, however. We
have tested the robustness of our procedure using different 
step sizes, ranging from $\sim 0.1 \;h^{-1}$ Mpc to 
$\sim 0.3 \;h^{-1}$ Mpc to find that the former is too 
small, since it will only locate the highest density peaks in the galaxy or
hot gas distribution, while the latter is somewhat too large since it 
typically registers very low density fluctuations, comparable to
the background level of the cluster image. 
The above
spatially defined procedure overcomes the difficulty of determining the density
thresholds in the two intrinsically different (optical and X--ray) 
density distribution. 

Note that for each cluster we find the highest cluster density-peak 
$(x_{\rm p}, y_{\rm p})$ within a radius of $\sim 0.5 \;h^{-1}$ Mpc around 
the original APM cluster center.
We then redefine the cluster center as being 
$(x_{\rm p}, y_{\rm p})$ and estimate all shape parameters around this 
new coordinate center. Typically, this coincides with the registered APM
cluster center.
\begin{figure}
\mbox{\epsfxsize=8cm \epsffile{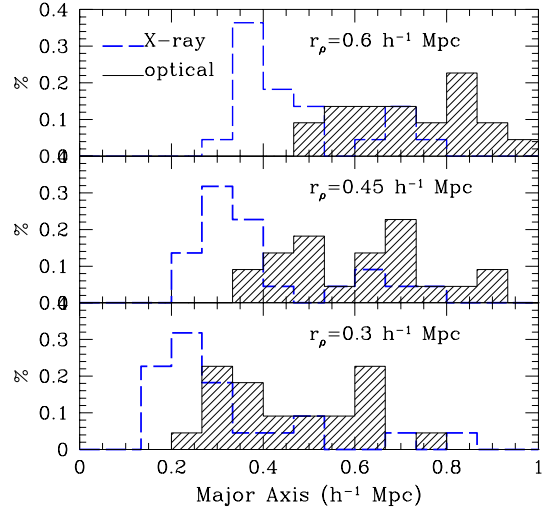}} 
\caption{Distribution of major axes of the fitted ellipses
for the optical (hatched) and X-ray cluster images and for the three density
thresholds based on the indicated $r_{\rho}$ (see definition in the text).}
\vspace{-0.5cm}
\end{figure}
We can get an idea of the regions of the clusters sampled by our procedure by 
inspecting Figure 1 where we plot the resulting major axis of the fitted
ellipses in each distribution and for the three density thresholds. It is 
evident that the regions sampled in the X-ray images are more centrally 
concentrated, as expected.

\subsection{Substructure Measures}

\subsubsection{Ellipticity}
It is expected that the existence of significant substructure affects the 
shape of the cluster in the direction of producing large ellipticities 
(McMillan et al. 1989; Davis \& Mushotzky 1993; West et al. 1995; Gomez et 
al. 1997). Although this appears to be a robust prediction, a small 
$\epsilon$ does not always endorse the lack of substructure. The reasons 
being that (a) small scale structure could develop symmetrically around the 
cluster core, and (b) a possible merger could happen along the line of 
sight. In both cases a small ellipticity would be measured.

\subsubsection{Cluster centroid shift}
Evrard et al. (1993) and Mohr et al. (1993) have suggested as an 
indicator of cluster substructure the shift of the center-of-mass 
position as a function of density threshold above which 
it is estimated (see also CER96; T98). Following their suggestion, 
we define as {\em centroid-shift} ($sc$)  the distance between the cluster 
center-of-mass, $(x_{\rm o}, y_{\rm o})$, where $x_{\rm o}=\sum\,x_{\rm i}\,
\rho_{\rm i}/\sum\,\rho_{\rm i}$, $y_{\rm o}=\sum\,y_{\rm i}\,\rho_
{\rm i}/\sum\,\rho_{\rm i}$ and the highest cluster density-peak, ie., 
\begin{equation}\label{eq:sc}
sc = \sqrt{(x_{\rm o}-x_{\rm p})^{2}\,+\,(x_{\rm o}-x_{\rm p})^{2}}\,.
\end{equation}
Notice here, that while the cluster center-of-mass changes as a function of 
density threshold ($\rho_{\rm t}$), above which we define 
the cluster shape parameters, the
position $(x_{\rm p}, y_{\rm p})$ remains unchanged. A large value of $sc$ may
therefore furnish a first clear indication of substructure.

In order to quantify the significance of such centroid variations to
the  presence of background contamination and random density
fluctuations, we carry out, in a fashion similar to BPM, a series of
Monte Carlo cluster simulations in which we have, by construction, no
substructure.  For each cluster we produce a series of simulated
clusters having the same number of observed galaxies as well as 
a random distribution of background galaxies, determined by the
distance of the cluster and the APM selection function. Furthermore,
the simulated galaxy distribution follows a King-like profile:

\begin{equation}\label{eq:sb}
\Sigma(r) \propto \left[1\,+\,\left(\frac{r}{r_{\rm c}} \right)^{2}
\right]^{-\alpha} \;,
\end{equation}

\noindent 
where $r_{\rm c}$ is the core radius, $\alpha = (3 \beta -1)/2$ and $\beta$ 
being the ratio of the specific energy in the galaxies to the specific thermal 
energy in the gas. We use the weighted, by the sample size, mean of 
most recent $r_c$ and $\alpha$ determinations 
(cf. Girardi et al. 1995; 1998), i.e., $r_{\rm c}=0.085 \;h^{-1}$ Mpc
and $\alpha=0.7$. 
We do test the robustness of our results for a plausible range of these
parameters. In general we find that the significance of the $sc$ measure 
decreases as $r_c$ increases. This is to 
be expected since using a large value of $r_c$, for the same number of 
core galaxies, will increase the random density fluctuations and thus
the $sc$ measure. Naturally, we expect our simulated clusters to generate
small $sc$'s and in any case insignificant shifts.
Therefore, for each optical cluster in our sample 
we perform 1000 such simulations and we derive $\langle sc \rangle_{\rm sim}$
as a function of the same thresholds, $\rho_{\rm t}$, 
as in the real cluster case. Then, within a search radius of
$0.75 \;h^{-1}$ Mpc from the simulated highest cluster peak,
we calculate the quantity:
\begin{equation}\label{eq:sig}
\sigma =\frac{\langle sc \rangle_{\rm o} - \langle sc 
\rangle_{\rm sim}}{\sigma_{\rm sim}}\;,
\end{equation}
in order to measure the significance of real centroid shifts
as compared to those of relaxed, mock objects. Note that 
$\langle sc \rangle_{\rm o}$
is the average, over the three density thresholds, centroid shift
for the real cluster. 

\subsubsection{Subgroup classes}
We also utilise a friend-of-friends algorithm to categorise the observed
substructure (see also section 3.2 of Rhee et al. 1991 for details). We join 
all cells having common boundaries that fall above each density threshold. 
We therefore create and register all subgroups as a function of 
$\rho_{\rm t}$ and rank substructure events according to the following 3 
categories (see BPM and Plionis et al. 2000 {\em in preparation}):  

(1) {\em No substructure}: Clusters with only one group at all density 
thresholds (regular, spherical systems). Systems with two unequal subgroups, 
where the second is much smaller than the first ($mincir 10\%$) 
also fall in this level. 

(2) {\em Weak or moderate substructure}: Objects with one clump at the first
$\rho_{\rm t}$ and multiple clumps at the next two levels (second or third 
threshold), where the second in size group is greater than 10\% and 
$\mincir$25\% of the total cluster size (see also RLT92). 

(3) {\em Strong substructure}: Like in (2) but now  the second in
size subgroup is $\magcir$25\% of  the total cluster size. Complex
systems with multiple condensations at all 
density thresholds are naturally included in this category. 
Finally, a cluster with two large 
clumps (bimodal) at the highest density level, 
of which one is $> 40\%$ of the 
other, fall in this category as well.

Note that in order to be consistent with the scale within which APM clusters 
are reliably identified and constructed (Dalton et al. 1997), we have used a 
radius of $0.75 \;h^{-1}$ Mpc as our maximum searching radius within which 
we search for subgroup statistics. We have also carried out the same analysis 
by increasing the radius to $1 \;h^{-1}$ Mpc but results do not change 
appreciably.
\begin{figure*}
\mbox{\epsfxsize=23cm \epsfysize=23cm \epsffile{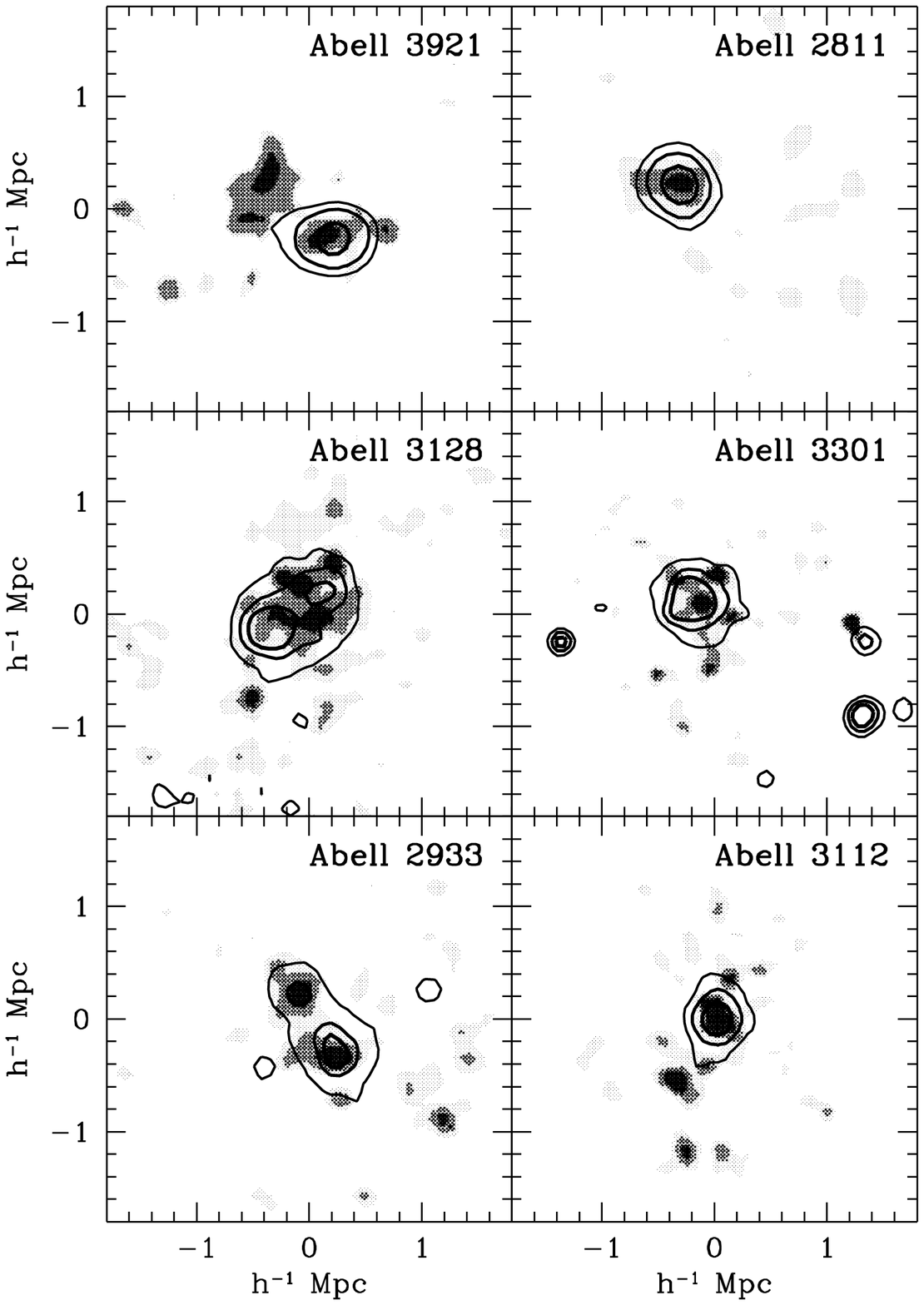}}
\end{figure*}
\begin{figure*}
\mbox{\epsfxsize=23cm \epsfysize=23cm \epsffile{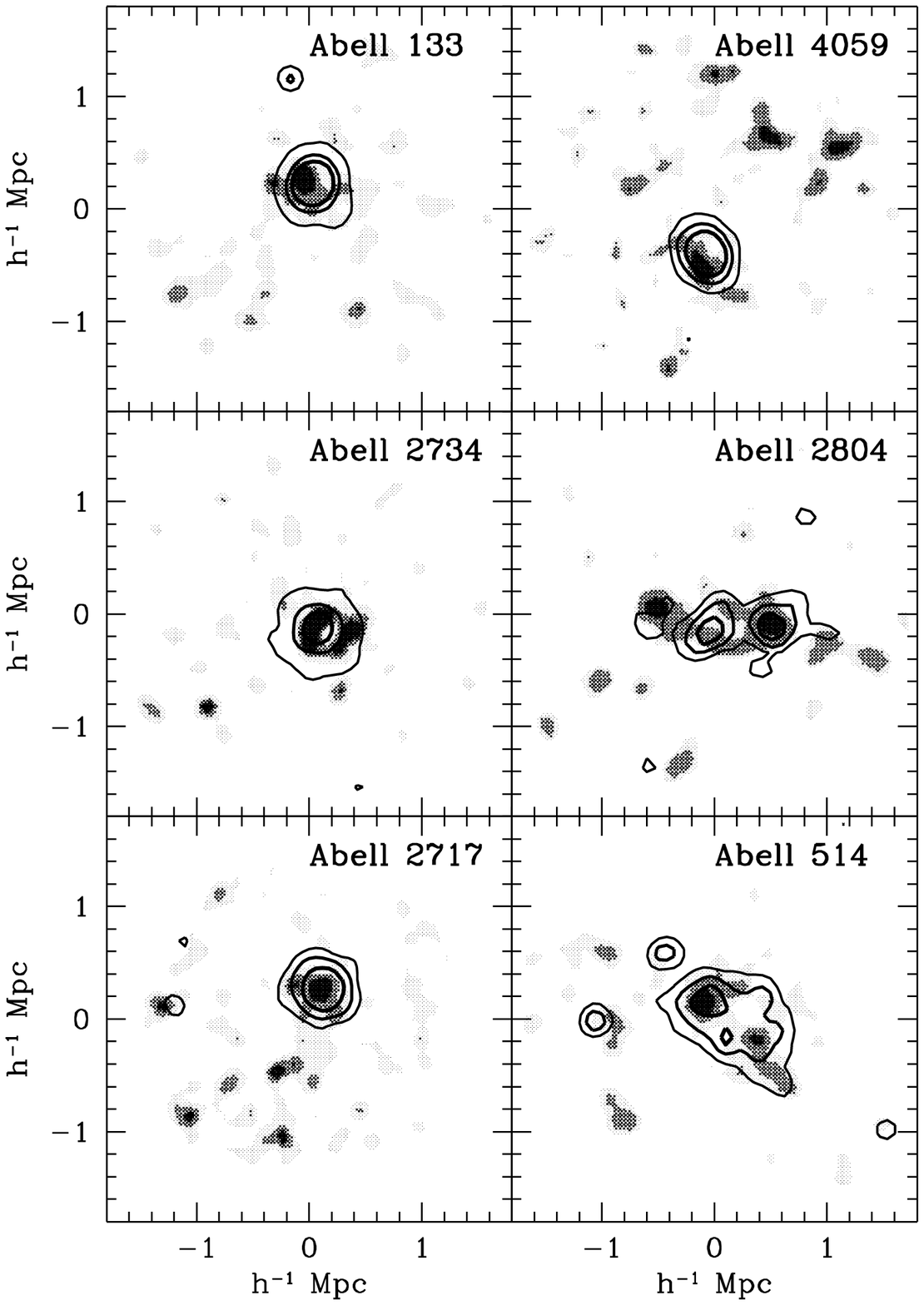}}
\end{figure*}
\begin{figure*}
\mbox{\epsfxsize=23cm \epsfysize=23cm \epsffile{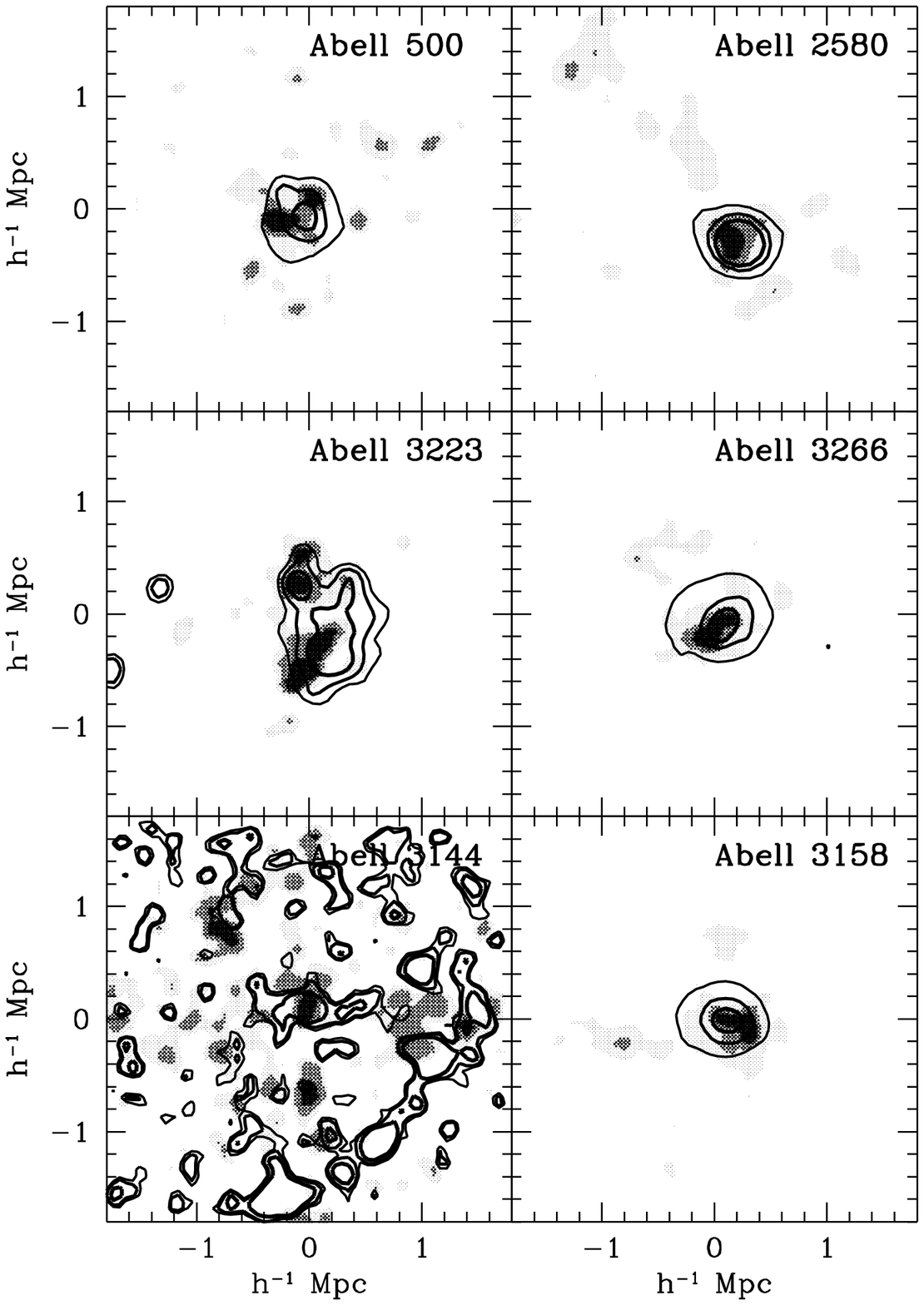}}
\end{figure*}
\begin{figure*}
\mbox{\epsfxsize=15.5cm \epsfysize=15.5cm \epsffile{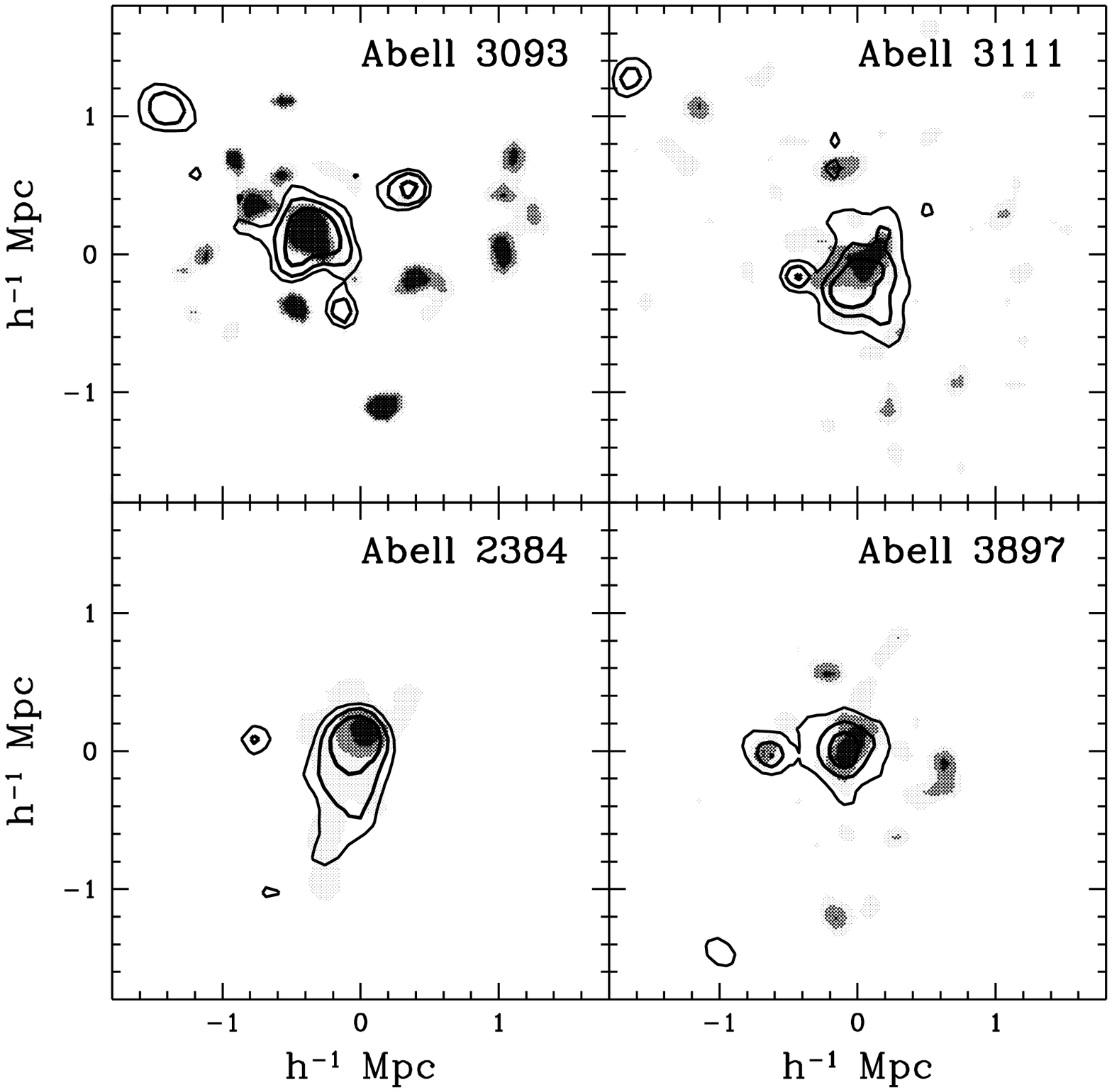}} 
\caption{Optical and X--ray images for our 22 galaxy clusters. Contours 
correspond to the X--ray images, whereas greyscale configurations denote 
the optical data. All axes are  expressed in physical units, while the 
scale of each image is $\sim 3.6 \;h^{-1}$ Mpc aside. North is top and East 
is right.}
\end{figure*}
 
\section{Comparison of Optical \& X--ray Cluster Images}
We investigate here the compatibility between the X--ray and optical cluster
data by visually inspecting their smoothed density distributions as well as 
by correlating their respective shape parameters (ellipticity and position 
angles). The aim of our procedure is to evaluate how well the optical APM 
data trace the cluster potential and therefore how reliable the optical 
data can be in deriving the structural and dynamical parameters of clusters. 
This is of paramount importance since large cluster samples exist mostly in 
the optical and their analysis can provide important constraints in theories 
of galaxy formation. 

A similar comparsion of optical and X-ray cluster data 
has been performed by Miller, Melott \&
Nichol (2000), and they found that optical total cluster luminosity is
directly proportional to the total X-ray luminosity for several orders
of magnitude in mass and luminosity (from poor groups to rich
clusters). These results point in the direction that data from both parts of
the spectrum can uniquely characterise the dynamical state of a cluster.

\subsection{Isodensity maps}

  We plot in Figure 2 the smooth APM galaxy distribution as greyscale
  maps and the X--ray data as isophote contour maps.  The three X--ray
  contours correspond to the density thresholds defined
  in section 3.2. As expected, clusters in
  the X--ray  appear to be more concentrated around the central
  potential wells, thus  having rounder contours than their optical
  counterparts. Instead, optical data depict more distinct structure
  (in the form of groups of galaxies) around the core, although this
  could in some cases, be due to projection effects.  A careful
  comparison with the X--ray maps shows that a few of the
  clusters in the optical may suffer from such problems, although in
  some cases galaxy groups,  visible only in the optical, may be weak
  X--ray emitters and hence absent from the X--ray
  images. Nevertheless, the majority of our clusters seem to have a
  nice agreement in both parts of the spectrum. In 17 out of 22
  systems, the gross features (double and secondary components,
  elliptical structures, multimodal objects  and single, relaxed
  configurations) of the mass distribution are apparent in both
  images. This is an indication that galaxies and groups of galaxies
  do trace the hot gas distribution in most of the cases ($\sim 80\%$). 
  In only five systems (A2717, A3112, A3897, A3093, A3921),
  we observe significant apparent substructure in the optical which is
  (almost) undetectable in the X--rays.

In some of the 17 clusters, with relatively good optical and 
X--ray image correspondence, there is evidence of recent merger events.
Zabludoff \& Zaritsky (1995) 
and  Baier et al. (1996) suggest that 
a substantial spatial difference between the optical
and the X--ray peak positions together with the X--ray peak being
distorted in an orthogonal direction with respect to the line
connecting the two main optical cluster clumps, signify two
undeniable collision vestiges. The reason is that gas, unlike galaxies
and DM, is collisional. This has been born out also from simulations
(Evrard 1990; Roettiger et al 1993) which have shown that 
during the collision of two groups,
the gas is stripped and remains in dis-equilibrium for $\sim 1 \;
h^{-1}$ Gyr afterwards. Therefore, the
criterion of a spatial displacement between the optical
and X--ray peak positions
seems to be particularly sensitive to the value of
$\Omega_{\circ}$, since it overcomes the known uncertainty of the cluster
relaxation time which hampers the cosmological interpretation of the
existence of cluster substructure.
Such indications, with varying strength, are apparent in A3128, A2804, 
A3223 and A500 where the corresponding differences
between optical and X--ray cluster peaks ($dp$) are of order $\sim
0.42 \;h^{-1}$ Mpc, $\sim 0.58 \;h^{-1}$ Mpc, $\sim 0.62
\;h^{-1}$ Mpc and $\sim 0.3 \; h^{-1}$ Mpc respectively. 
The argument of orthogonality suits best A3128 as is further evident from the 
large misalignment angles at all $\rho_{\rm t}$'s between the optical and the 
X--ray mass distributions.
\begin{figure}
\mbox{\epsfxsize=7cm \epsffile{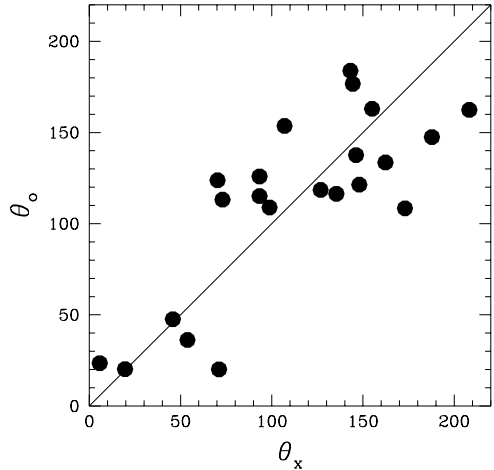}} 
\caption{Comparison of the X--ray (ROSAT) and optically (APM) defined cluster 
position angles. The open symbols correspond to the three clusters that have 
discordant position angles (see text). The solid line corresponds to a 
perfect correlation.}
\end{figure}
\subsection{Position angles}
As a first quantitative test of the compatibility between the X--ray and 
optical images we correlate in Figure 3 their respective 
cluster major axis orientations. It is evident that there
is a good correlation, with coefficient $\sim 0.8$ and probability
of no correlation ${\cal P} \simeq  10^{-5}$.
No cluster has $\delta \theta \magcir 60^{\circ}$, while the mean 
misalignment angle between the cluster optical and X--ray 
defined position angles is 
$\langle \delta \theta \rangle \mincir 28^{\circ} \pm 18^{\circ}$.
Excluding the five clusters that we have identified as
having discordant optical and X--ray morphological features we find
a similar correlation coefficient but with a slightly
lower significance  (${\cal P} \simeq 10^{-4}$) which
is mostly due to the reduction of the sample.
\begin{figure}
\mbox{\epsfxsize=7cm \epsffile{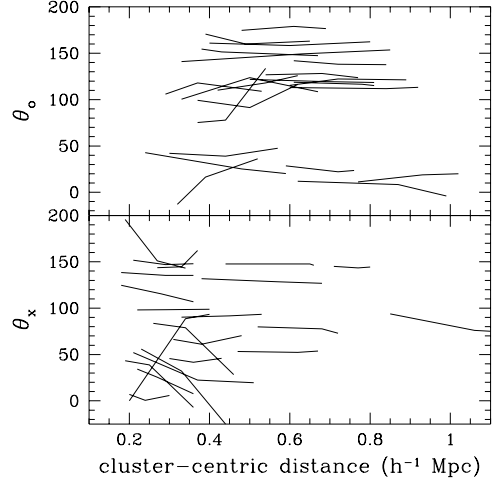}} 
\caption{Position angle variation as a function of cluster-centric distance
for all 22 clusters.}
\end{figure}

Davis \& Mushotzky (1993) have argued that considerable variations of 
cluster orientations as a function of distance from the cluster center 
could be considered as evidence of merger events.  If, 
however, subgroups dominantly develop along the filamentary 
structure in which the cluster is embedded then variations in the 
cluster position angle could be negligible and thus 
significant subclumping can escape detection (see West 1995; 
West et al. 1995). In Figure 4 we examine possible variations in cluster 
orientations as a function of $\rho_{\rm t}$ and thus as a function of 
distance from the cluster center. For each cluster we connect the 
position angles estimated at the three density thresholds, and thus at three 
different distances from the cluster center. It is evident that 
only a few clusters exhibit evidence for such an effect in their X-ray 
images, which however should be attributed to their weak position angle 
determination (due to their extremely small ellipticity).

\subsection{Ellipticities} 
We calculate cluster ellipticities in both data sets as a
function of $\rho_{\rm t}$. The cluster $\epsilon$'s
defined in X--rays are generally slightly smaller than their optical
counterparts, corresponding to more spherical configurations (especially 
for single-component clusters). This is expected when the
inter-cluster gas is in hydrostatic equilibrium and the dark matter 
is distributed like galaxies. However, in the case where we have
evidence of recent 
merger events (A3128, A2804, A3223) we observe that the X-ray ellipticity 
is typically larger than the corresponding optical. Due to these 
conflicting trends the correlation between optical and X-ray ellitpicities 
is rather weak as can be seen in Figure 5. Excluding the five systems with 
discordant optical and X--ray morphologies (A2717, A3112, A3897, A3093, 
A3921) and A3128, in which due to a recent merger there is strong 
differentiation of its X-ray and optical morphology, we find a correlation 
coefficient of $\sim 0.7$ with the probability of zero correlation being 
${\cal P} \simeq 3 \times 10^{-3}$. 

\begin{figure}
\mbox{\epsfxsize=7cm \epsffile{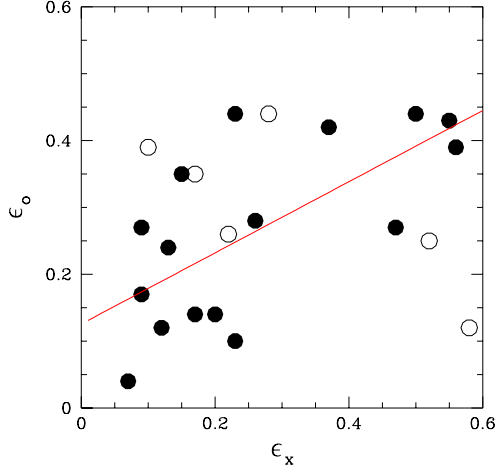}} 
\caption{Comparison of optical and X--ray defined cluster $\epsilon$'s. The 
5 clusters with discordant morphologies and A3128 are shown as open circles.}
\end{figure}

Another characteristic is that for most clusters (17/22) their optical
$\epsilon$ is a weakly decreasing 
function of cluster-centric distance, as can be seen in Figure 6, which
should be attributed mostly to the effect of background galaxies projected in
the area covered by the cluster. In the X-ray images such an effect is
evident only in 9 clusters, most of which however are those having
large ellipticity. Visual inspection of these clusters show that the
effect is not artifiscial and should be attributed to bimodality at the
highest threshold or to the existance of
significant substructure located in the central parts of the clusters
(cf. A3128, A3223, A2933 etc). 

\begin{figure}
\mbox{\epsfxsize=7cm \epsffile{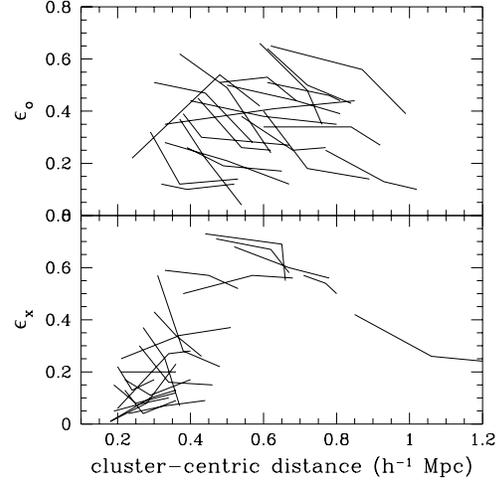}} 
\caption{Ellipticity variation as a function of cluster-centric distance
for all 22 clusters.}
\end{figure}

For A2717, A3112, A3897 and A3921 their large $\epsilon_{\rm o}$ does not
correspond to a similarly large $\epsilon_{\rm x}$, supporting the view
that these clusters may suffer from optical projection effects that have
altered their true 2D structure (see also section 4.1). 

A particularly interesting example is the noisy and apparently highly unrelaxed
A3144 cluster. However, its relatively small value of $\epsilon$ could be 
possibly attributed to the development of small-scale structures 
symmetrically around the cluster core or to the existance of
significant noise.

\section{ Substructure Results}
We present in this part the results of all the substructure tests that we have
applied to the optical and X-ray cluster data. The results are tabulated in 
Table 2, together with the cluster shape parameters, discussed in the 
previous section.

\subsection{Centroid shifts}
We calculate $sc$, ie., the difference between the position of the weighted
cluster center and the highest density peak as a function of
$\rho_{\rm t}$ within a radius of $\sim 0.75 \;h^{-1}$ Mpc (equation 5). 
In Figure 7 we correlate the optical and X--ray estimated centroid shifts. 
We observe a very good correlation between the relative cluster centroid 
variations for the 17 clusters that have concordant morphological features 
in both optical and X--rays data (correlation coefficient $\sim 0.8$ and 
${\cal P} \simeq 3 \times 10^{-4}$). As expected the significance of the 
optical $sc$ (equation 7) is well correlated with the value of the optical and 
X-ray $sc$ ($R\sim 0.9$ and $\sim 0.8$, respectively).

\begin{figure}
\mbox{\epsfxsize=7cm \epsffile{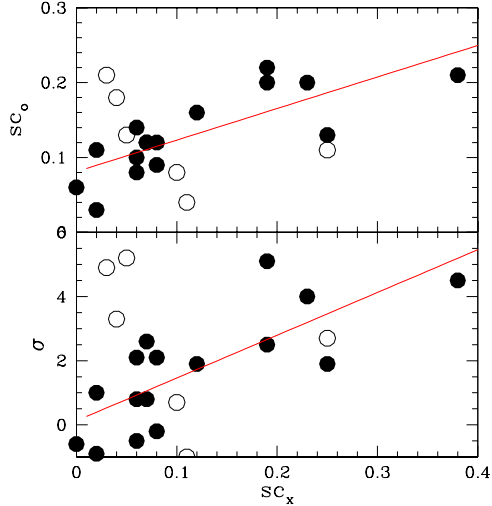}} 
\caption{Centroid shift statistics. Upper panel: 
Optical versus X--ray centroid shift values. Lower panel: X-ray $sc$ values 
versus optical $sc$ significance. With open circles we depict the 6 clusters 
with discordant optical \& X--ray morphological features.}
\end{figure}

{\em These results indicate that the $sc$ measure and its 
significance is a useful tool to characterize the degree 
of cluster substructure using optical or X-ray data.} 

Most of the five clusters that are suspect of being affected by projection
effects in the optical (A2717, A3112, A3921) have a large $sc$ value 
(only in the optical) and correspondingly high significance indication of 
subclustering. This fact further suggests that these clusters suffer from 
projection effects. The cases of A3897 and A3093 are somewhat different, as 
it is evident from Figure 2.  Although there is evidence of projection 
effects it appears that substructure emerges in a symmetric fashion around 
the optical cluster core. Such an effect is more pronounced in the optical 
than in the X--ray images, thus producing as expected a small $sc$ value.  


\subsection{Subgroup statistics}
Using the categories defined in section 3.3.3 and also a maximum search 
radius of $\simeq 0.75 \;h^{-1}$ Mpc, we obtain 6 clusters falling into the 
first category of relaxed objects, 6 clusters in the category of systems 
displaying partial (weak) substructure and 10 clusters belonging to the third 
case within which clusters exhibit obvious and concrete indications of 
subclumping events. Note that 4 clusters of the strong substructure case are 
the ones affected by projections (A3093, A3112, A3897, A3921). In the last 
column of Table 2 we show the index corresponding to each cluster 
subgroup index.

The results of this analysis are in good broad agreement with the other 
substructure measures ($\epsilon$'s and $sc$'s), with differences only in a 
few cases. The vast majority of clusters (19 out of 22) show that their 
subgroup index does agree, at least qualitatively, with the 
moments analysis that we have applied in this work (section 3). Nevertheless, in 3 
cases (A3266, A3301, A2384) results imply discrepancies between our 
percolation-like algorithm and the method of moments. On checking the 
parameter space of both techniques, we have found that while the latter stays 
firm, the former is extremely sensitive to slight alterations regarding the 
geometric definition of our classes (section 3.3.3). 

In the line of the above arguments, we suggest that it is wise to use these 
results on an advisory level and not as definite. 
   
\begin{figure}
\mbox{\epsfxsize=7cm \epsffile{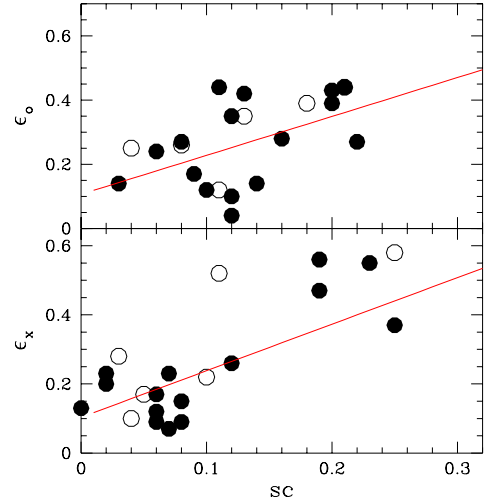}} 
\caption{Cluster ellipticities as a function of centroid shifts in the 
optical (top panel) and X--ray (bottom) data, together with the lines of 
the best fit. Open symbols are as in Figure 7.}
\end{figure}

\begin{table}
\caption[]{Results on cluster substructure.  Columns 3 to 5 are 
the $sc$, $\epsilon$ and $\theta$ in
the X-ray, columns 6 to 8 are the analogous parameters in the
optical data, column 9 shows the significance of $sc_{\rm o}$, estimated
using equation \ref{eq:sig} and column 10 refers to the subgroup 
category (see section 3.3.3).
Note that centroid variations are given in $h^{-1}$ Mpc and 
position angles in degrees.}  
\tabcolsep 3pt
\begin{tabular}{cccccccccc} \hline 
ACO & APM & $sc_{\rm x}$ & $\epsilon_{\rm x}$ & $\theta_{\rm x}$ &
 $sc_{\rm o}$ & $\epsilon_{\rm o}$ & $\theta_{\rm o}$ & $\sigma$ & PG
\\ \\ \hline \hline
2717&   5 & .04 &.10 &143.1 & .18 &.39 &  5.1 & 3.3 & 1 \\
2734&  15 & .06 &.09 &  7.8 & .08 &.27 &147.5 & 0.8 & 2 \\
2804&  99 & .19 &.56 & 73.0 & .22 &.27 &113.2 & 2.5 & 2 \\
2811& 104 & .00 &.13 &135.4 & .06 &.24 &116.4 &-0.6 & 1 \\
0133& 138 & .06 &.12 &173.2 & .10 &.12 &108.4 &-0.5 & 2 \\
2933& 204 & .23 &.55 &146.2 & .20 &.43 &137.6 & 4.0 & 3 \\
3093& 347 & .10 &.22 & 70.3 & .08 &.26 &123.8 & 0.7 & 3 \\
3111& 373 & .08 &.15 & 28.3 & .12 &.35 &162.4 & 2.1 & 2 \\
3112& 374 & .05 &.17 &  5.7 & .13 &.35 & 23.5 & 5.2 & 3 \\
3128& 403 & .25 &.58 & 53.8 & .11 &.12 & 36.2 & 2.7 & 3 \\
3144& 427 & .07 &.23 & 71.1 & .12 &.10 & 20.1 & 2.6 & 3 \\
3158& 434 & .06 &.17 & 98.8 & .14 &.14 &108.9 & 2.1 & 1 \\
3223& 484 & .38 &.50 &144.4 & .21 &.44 &176.7 & 4.5 & 3 \\
3266& 510 & .12 &.26 & 45.8 & .16 &.28 & 47.6 & 2.0 & 1 \\
0500& 518 & .07 &.07 &162.3 & .12 &.04 &133.6 & 0.8 & 2 \\
0514& 533 & .19 &.56 &126.8 & .20 &.39 &118.5 & 5.1 & 3 \\
3301& 560 & .08 &.09 &155.2 & .09 &.17 &163.0 & -.2 & 3 \\
2384& 725 & .25 &.37 & 19.6 & .13 &.42 & 20.2 & 2.0 & 1 \\
3897& 806 & .11 &.52 & 93.3 & .04 &.25 &125.9 &-1.0 & 3 \\
3921& 822 & .03 &.28 & 93.3 & .21 &.44 &115.1 & 4.9 & 3 \\
2580& 888 & .02 &.23 &107.0 & .11 &.44 &153.6 & 1.0 & 1 \\
4059& 938 & .02 &.20 &148.0 & .03 &.14 &121.4 &-0.9 & 2 \\ \hline
\end{tabular}
\end{table}

\subsection{Comparison of different substructure test}
In Figure 8 we correlate ellipticities and $sc$'s for all 22 clusters
of our sample. The upper panel
corresponds to the optical and the lower to the X--ray data. 
It is evident that the two substructure measures
are correlated in both sets of data. 
The correlation coefficients are $\sim 0.8$ and $\sim 0.5$ for the X--ray and
optical data respectively with probability of zero correlation
being ${\cal P} \simeq 10^{-4}$ for the X--ray 
and ${\cal P} \simeq 0.05$ for the optical data respectively.

Furthermore, cross-correlating, in figure 9, the optical $sc$'s and 
X--ray ellipticities 
we also find a significant correlation with ${\cal P} \mincir 10^{-4}$ and 
$R\sim 0.82$ if we exclude the clusters with discordant morphologies. 
We find that the line of best fit is given by: 
$$\epsilon_{\rm x} \simeq 2.55 (\pm 0.47) sc_{\rm o} + 0.07 (\pm 0.06)\;\;,$$ 
from which we can deduce the shape of the ICM region emitting X-rays,
directly from optical cluster data.

\begin{figure}
\mbox{\epsfxsize=8cm \epsffile{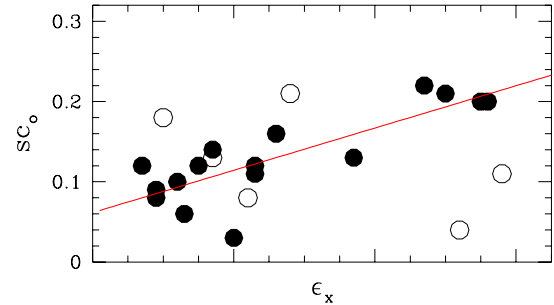}}
\caption{Cross-correlation between the cluster 
X--ray ellipticity and its optical $sc$ value. 
Open symbols denote the clusters 
with discordant optical and X--ray morphological parameters.} 
\end{figure}

These results imply that the flattening of clusters, as evident in 
X--rays as well as in the optical, is a result of their
dynamical activity and not due to initial conditions; for example
even high-peaks of an underlying Gaussian random field are aspherical,
although less so than lower density peaks (cf. Bardeen et al. 1986).


\subsection{Main Results}
From the total of 22 clusters, at least in 8 cases we observe strong 
substructure signatures, verified by all available methods and data (A2804, 
A2933, A3128, A3144, A3223, A3266, A514, A2384). Of the remaining population 
we have confirmed, using all our substructure indications, that 9 clusters 
(A2734, A2811, A133, A3111, A500, A3158, A3301, A2580, A4059) show no or 
weak substructure activity in both parts of the spectrum (although there is 
some evidence of a recent merger event in A500).  We have also found that 5 
clusters (A2717, A3112, A3921, A3093, A3897) appear distinctly bimodal (or 
even multimodal) although their X--ray contour maps are almost relaxed, which 
we attribute to projection effects in the optical. From these, A2717, A3112 
and A3093 seem to be single-component (relaxed) clusters (based on their 
X-ray images), while A3921 and A3897 shows clear indications of elliptical 
systems. Therefore we find that $\magcir 45\%$ of clusters exhibit 
significant evidence of substructure, which is in good general accordance 
with most substructure analysis results (see references in Introduction). 

Note that out of our 22 clusters, 11 (50\%) have been examined for 
substructure signals elsewhere in the literature (A2717, A133, A3158, A3128, 
A3266, A500, A514, A2384, A3897, A3921, A4059). We find that our results are 
in general good qualitative and quantitative agreement with those of other 
studies (see Appendix). Although, mostly different methods have been employed 
in these studies (wavelet transform analysis, isophotal maps, power ratios, 
kinematical and velocity dispersion estimators), we do not detect any serious 
disparities regarding the dynamical and shape parameters for these X--ray 
clusters that we have in common. For example we find our position angles 
always within $\approx 10^{\circ}$ of the other determinations.

\subsection{Morphological and dynamical classification}
We can now proceed in classifying the present sample according to a scheme 
that is close to the one developed by JF99 (see also Forman \& Jones 1990; 
Jones \& Forman 1992; Forman \& Jones 1994 and Girardi et al.  1997 hereafter 
G97). Just to be consistent with the largest and most complete of the above 
analyses (JF99), we will restrict ourselves only to the X--ray cluster 
images, which we regard as more reliable in terms of the current assessment 
(see Table 3).

Clusters with no or marginal evidence of subclumping events are dubbed as 
single, relaxed objects and given the symbol {\sc U}, featuring unimodality. 
In this category we count 12 X--ray clusters. Looking at the strong 
substructure cases, we have 3 bimodal ({\sc B}) objects (A2804, A2933 and 
A3128) out of which the first two show their respective optical doubles, 
while optical A3128 appears to be somewhat closer to a complex system. We 
characterise complex or multimodal ({\sc M}) systems that display more than 
two clumps in their contour maps. Typical such examples are A3144, A3223 and 
A514, also authenticated by inspection in the optical. Objects showing 
apparent deviations from singularity also having large $\epsilon$'s, without 
depicting obvious small-scale structures, are flagged as elliptical (A3266, 
A2384 and A3921). The latter three clusters, in absence of any other 
substructure characteristic, are tagged as {\sc E}'s. Finally, A3897 is 
tagged as {\sc P}, characteristic of a large central region (primary 
component) associated with a small secondary structure (left of the X--ray 
image), visible in both data. Note here, that its optical counterpart would 
have been flagged as complex, since it shows two extra small-scale clumps at 
the top and right of the contour plot (see Figure 2).

\begin{table}
\caption[]{Codification of our sample according to a modified JF99 
classification scheme.  The first column denotes the type of activity, 
second shows the number of clusters falling into that category, third 
presents its respective fraction to be compared with JF99 results in 
column 4. Note that the maximum 
scale of detecting cluster substructure in JF99 is $0.5\;h^{-1}\;$Mpc, so 
that the following comparison should be seen with caution.}  
\tabcolsep 11pt
\begin{tabular}{cccc}
 \hline  Class & Number & Fraction  & JF99    \\ \hline \hline
  {\sc U}   &   12   &   0.550    &  0.560   \\ 
  {\sc B}   &    3   &   0.135    &  0.060 \\
  {\sc M}   &    3   &   0.135    &  0.130  \\
  {\sc E}   &    3   &   0.135    &  0.140  \\
  {\sc P}   &    1   &   0.045    &  0.030 \\ \hline
\end{tabular}
\vspace{-0.2cm}
\end{table}

\section{Concluding remarks}

We have investigated a sample of 22 galaxy clusters using optical (APM) and 
X--ray (ROSAT) data with the aim of addressing two questions: (a) Do optical 
and X--ray data reveal the same cluster morphological features ? (b) What is 
the percentage of relaxed and dynamically active clusters ?

Our cluster sample has flown from a cross-correlation between the APM, ACO 
and ROSAT pointed observations data and has a depth distribution with 
$z \mincir 0.13$. We have examined our cluster sample utilising several 
cluster-morphology diagnostics such as isodensity maps, orientations, 
ellipticities, centroid variations and subgroup statistics. Looking at the 
isodensity contour maps, the cluster orientations and ellipticities, we 
observe a remarkable 1-to-1 correspondence between X--ray and optical data 
in $\sim 80\%$ of our sample, regarding the gross cluster characteristics 
(prime structures, elongations, multimodality, collision signatures). We 
quantify this by correlating the optical and X--ray cluster ellipticities 
and orientations and find high and statistically significant correlations. 
In an attempt to quantify the compatibility of the different substructure 
measures, that we have used, we correlate $\epsilon$'s and $sc$'s, in the 
optical and in the X--ray separately and we also cross-correlate them to find 
significant and strong correlations. This implies that indeed the flattening 
of clusters is due to their dynamical activity.

From our substructure analysis we find that $\sim 10$ out of 22 systems 
($\sim 45\%$) display strong substructure indications visible in both parts 
of the spectrum. We also find that 5 clusters ($\sim 22\%$) show clear 
disparities between the optical and X--ray maps, with apparent substructure 
in the optical not corroborated by the available X--ray data. This is most 
possibly due to optical projection effects. Our results on the frequency of 
disordered clusters do concur with most of the relevant studies published 
to-date. 

On the understanding that our catalogue is rather inadequate for 
drawing significant cosmological conclusions, we would prefer to be cautious 
when it comes to such a precarious task. Nevertheless, we observe that our 
present analysis is compatible with that of RLT92 (see their Figure 2) 
regarding the cluster substructure frequency, setting a rather frail lower 
limit on the density parameter ($\Omega_{\circ}\,\geq\,0.5$; see also West 
1995; West et al. 1995). Furthermore, the relatively large fraction
(4/22) of recent mergers that we have identified is again indicative of
a high-$\Omega_{\circ}$ Universe (see discussion of Zabludoff 
\& Zaritsky 1995).

In the near future we plan to apply the methodology of this work to 
$\sim 900$ APM galaxy clusters, in order to 
investigate in more detail the issue of cluster substructure. 

\section*{Acknowledgements}
Both S. Basilakos and V. Kolokotronis acknowledge financial support from the 
Greek State Fellowship Foundation. M. Plionis acknowledges the hospitality
of the Astrophysics Group of Imperial College, were this work was completed.
This research work has made use of NASA 
Extragalactic Database (NED). The cluster data have been obtained through 
LEDAS online service, provided by the University of Leicester.

\appendix
\section{Details on individual clusters}

${\bf A2717}$: Obvious similarity of the primary cluster peak ($\leq 0.4
\;h^{-1}$ Mpc) but disparity on the secondary structure which is only 
distinct in the optical image. Suspect of being affected by projection effects.
Eight X--ray and  radio sources have been 
removed due to their point-like nature (see also Slezak et al. 1994; Mohr et 
al. 1995; BT96; G97). 

\noindent ${\bf A2734}$: Apparently good correspondence between the two 
images. Two radio point-like sources (center and upper left) have been 
subtracted from the X--ray image. 

\noindent ${\bf A2804}$: Highly elongated cluster with large and significant 
$sc$'s in both images. Collision signature visible since the 
primary optical cluster structure seems to be displaced with respect to its 
X--ray analogue by $0.58\;h^{-1}\;$Mpc. Two radio point-like sources have 
been excised from the X--ray image.

\noindent ${\bf A2811}$: A relaxed single-component system corroborated by 
both data, typical of a unimodal configuration. No sources have been removed 
here.

\noindent ${\bf A133}$: Probably a confirmed unimodal cluster by all 
substructure measures (Mohr et al. 1995; BT96) showing small values of 
ellipticities and $sc$'s. Optical and X--ray $\theta$'s differ by more than 
$64^{\circ}$. One projected X--ray source at $z\sim 0.235$ has been 
subtracted (EXO 0059.8-2218). 

\noindent ${\bf A2933}$: This is the archetype of a bimodal cluster as 
computed by both data. Highly significant $sc$'s, large 
ellipticities ($\simeq 0.5$) and similar $\theta$'s granting excellent 1-to-1 
correspondence. One point-like source has been removed from the upper right 
of the X--ray image.

\noindent ${\bf A3093}$: Four point-like sources have been excised
from the  X--ray data (center bottom and left). This system has
mediocre ellipticities and non-significant centroid shifts. 
It also yields a $\delta\theta\simeq 54^{\circ}$. 
Suspect of being affected by projection effects.

\noindent ${\bf A3111}$: Partial substructure activity present at vary low 
density thresholds. If taken at face value, system would have been flagged as 
complex. Isodensity contour plots are very similar, whereas its respective 
$\delta\theta$ is more than $45^{\circ}$. Like the previous one and in the 
absence of any distinct activity, it is regarded as a relaxed object. Five 
point-like sources have been removed from the X--ray cluster map.

\noindent ${\bf A3112}$: No sources were subtracted here. This is the 
archetype of image disparity. Obvious bimodality in the optical 
(a $5.2\sigma$ $sc$ event) corresponds to definite unimodality in the 
X--rays. Excluded from the cross-correlation statistical analysis (together 
with A2717, A3897, A3921 and A3093) as being suspect of optical projections. 

\noindent ${\bf A3128}$: One radio point-like source (PMN: JO331-S242) has 
been removed (lower right of X--ray map). Dissimilar cluster orientations and 
ellipticities are typical of a collision event between the prime cluster 
structures. Despite that, it exhibits highly significant $sc$'s ($\sim 
3\sigma$) in both data. In the G97 analysis this cluster is dubbed as a 
unimodal object on a scale of $\sim 1\;h^{-1}\;$ Mpc. Notice that the latter 
study is based on an entirely different approach than the one developed here. 

\noindent ${\bf A3144}$: This is a typical complex system. Multiple peaks 
associated with marginally significant $sc$'s and medium $\epsilon$'s due to 
symmetrically developed structures around the central cluster potential wells. 
No sources have been excluded here. 

\noindent ${\bf A3158}$: A single-component cluster, well-aligned and also 
showing small values of ellipticities and insignificant $sc$'s (Slezak et 
al. 1994; Mohr et al. 1995; BT96; G97). The optical $sc\sim 0.14\;h^{-1}
\;$Mpc is only a $2\sigma$ substructure event. No sources have been removed.

\noindent ${\bf A3223}$: Typical complex system displaying large $\epsilon 
\simeq 0.5$ and the largest $sc\sim 0.38 \;h^{-1}\;$Mpc out of the whole 
sample. Optical peak appears to be largely displaced with 
respect to the X--ray one ($dp\,\sim 0.62 \;h^{-1}\;$Mpc), 
a fact that signifies 
a collision vestige. Four radio point-like sources have been subtracted from 
the X--ray map (upper left, right and lower left).

\noindent ${\bf A3266}$: We have classified this cluster as elliptical in 
the absence of other distinct features. Optical and X-ray images
are well-aligned. No sources excised (see also Mohr et al. 1993; 
1995; BT96; G97; de Grandi \& Molendi 1999).

\noindent ${\bf A500}$: Good correspondence between optical and X--ray 
isodensity maps. It shows weak substructure which is rather insignificant. 
However, there are indications of a possible recent merger 
(displacement of optical and X-ray peaks by $dp\leq 0.3\;h^{-1}\;$Mpc).
No sources 
have been removed from the X--ray map. Similar studies (Mohr et al. 1995; 
BT96) show evidence of a unimodal cluster at least within 
$\sim 0.5\;h^{-1}\;$Mpc in all substructure properties.

\noindent ${\bf A514}$: Another classical multimodal system with 
distinct density peaks in both maps. Large ellipticities are followed 
by analogously large and statistically significant $sc$'s ($\sim 5\sigma$). 
Two X--ray and two radio point-like sources have been omitted here. This is 
the complex archetype in most of the published works up till now
(see also West et al. 1995; 
BT96; Bliton et al. 1998; JF99).

\noindent ${\bf A3301}$: Two point-like sources have been subtracted from the 
left of the X--ray map. Definitely unimodal in the X--ray but slightly 
elongated and multimodal in the optical, although with an insignificant $sc$. 
Relatively good 1-to-1 correspondence of the contour maps. 

\noindent ${\bf A2384}$: This is the archetype of an elliptical cluster. A 
high quality HRI image ($t_{\rm exp}>$7 hrs) which exhibits large $sc$'s  
and $\epsilon$'s in both images. However, the $sc$ values are marginally 
significant. No sources have been removed in this case (see also McMillan et 
al. 1989; West et al. 1995).

\noindent ${\bf A3897}$: An object with distinct substructure in both data, 
typical of the category {\sc P} (see X--ray map). Notwithstanding that, it 
seems multimodal in the optical. The optical $sc$ is non-significant but 
it is apparent that it is somehow 
underestimated due to symmetric and equally-sized structures developing 
around the central core. The same reasoning fully explains the low 
$\epsilon_{\rm o}$. Suspect of being affected by projection effects.
Three point-like sources have been excised at the lower 
left of the X--ray image (see also Gomez et al. 1997).

\noindent ${\bf A3921}$: A definitely optically complex object which is 
seemingly unimodal in the X--rays. There are, however, traces of elongation 
in the X--ray contour plot, only at the lower $\rho_{\rm t}$. This extension 
appears to be in the exact direction of the secondary optical structure which 
is not visible in the X--ray map. As a result, both data maps seem to be 
well-aligned and we have therefore classified A3921 as an elliptical object
but we have also considered it as being affected by projection effects in the 
optical. Three point-like sources (center and lower left) have been removed 
from the X--ray cluster (see also Mohr et al. 1995; BT96). 

\noindent ${\bf A2580}$: Within $0.5\;h^{-1}\;$ Mpc of the highest cluster 
peak, this cluster seems relaxed and unimodal in both images. There is some 
substructure evidence by means of the optical $sc$ (non-significant) and 
ellipticity, while it appears slightly misaligned 
($\delta\theta \leq 47^{\circ}$ ). No sources have been subtracted from this 
system.

\noindent ${\bf A4059}$: This is another typical unimodal system. 
A low $sc$ object which displays remarkable accordance in the isodensity 
maps and a $\delta\theta \leq 27^{\circ}$. No sources have been removed from 
this one. Definitely dubbed as a single-component cluster also by other 
analyses (cf. Slezak et al. 1994; Mohr et al. 1995; BT96; G97).

\end{document}